\documentclass[12pt,preprint,iop,apj]{emulateapj}  
\usepackage{graphicx}
\usepackage{txfonts}
\newcommand{\vai}{v_{\mathrm{A,i}}}

\newcommand{\der}{{\rm d}}

\newcommand{\rhoi}{\rho_{\rm i}}
\newcommand{\rhoe}{\rho_{\rm e}}

\newcommand{\rhotr}{\rho_{\rm tr}}

\newcommand{\ra}{r_{\mathrm{A}}}

\usepackage{natbib}

\begin{document}

	\title{The behavior of transverse waves in nonuniform solar flux tubes.\\ II. Implications for coronal loop seismology}

	\shorttitle{Transverse waves. II. Seismology implications}

  \author{Roberto Soler$^1$, Marcel Goossens$^2$, Jaume Terradas$^1$, and Ram\'on Oliver$^1$}
     \affil{$^1$Departament de F\'isica, Universitat de les Illes Balears,
               E-07122, Palma de Mallorca, Spain}
  \affil{$^2$Centre for Mathematical Plasma Astrophysics, Department of Mathematics, KU Leuven,
             Celestijnenlaan 200B, 3001 Leuven, Belgium}
              \email{roberto.soler@uib.es}

  \begin{abstract}

Seismology of coronal loops using observations of damped transverse oscillations in combination with results from theoretical models is a tool to indirectly infer physical parameters in the solar atmospheric plasma. Existing  seismology schemes based on  approximations to the period and damping time of kink oscillations are often used beyond their theoretical range of applicability. These approximations assume that the variation of density across the loop is confined to a nonuniform layer much thinner than the radius of the loop, but the results of the inversion problem often do not satisfy this preliminary hypothesis.  Here, we determine the accuracy of the analytic  approximations to the period and damping time, and its impact on seismology estimates, when largely nonuniform loops are considered. We find that the accuracy of the approximations when used beyond their  range of applicability is strongly affected by the form of the density profile across the loop, that is observationally unknown and so must be arbitrarily imposed as part of the theoretical model. The error associated with the analytic approximations can be   larger than 50\%  even for relatively thin nonuniform layers. This error directly affects the accuracy of  approximate  seismology estimates  compared to  actual numerical inversions. In addition, assuming different density profiles can produce noncoincident intervals of the seismic variables in  inversions of the same event. The ignorance about the true shape of density variation across the loop is an important source of error that may dispute the reliability of parameters seismically inferred  assuming an ad hoc density profile.

  \end{abstract}

   \keywords{Sun: oscillations ---
                Sun: atmosphere ---
		Sun: magnetic fields ---
		waves ---
		Magnetohydrodynamics (MHD)}

%________________________________________________________________

\section{INTRODUCTION}

Coronal seismology was first suggested by \citet{uchida1970} and \citet{rosenberg1970}, although the paper by \citet{roberts1984} is often credited as the work that set the foundations of the method. The coronal seismology technique combines observations of magnetohydrodynamic (MHD) waves in the solar corona with the predicted behavior of these waves in theoretical models in order to indirectly infer physical properties of the coronal plasma and magnetic field \citep[see various examples in, e.g.,][among others]{roberts1984,nakariakov2001,andries2005seis,verwichte2006,verth2008,verth2010,arregui2013prop}. A particular example of this method is the inversion of physical conditions in coronal loops using the observed period, $P$, and  damping time, $\tau_{\rm D}$, of their transverse oscillations  \citep[e.g.,][]{nakariakov1999,aschwanden1999,aschwanden2002,ofman2002} along with the theoretically predicted values based on an interpretation in terms of resonantly damped kink MHD waves \citep[e.g.,][]{rudermanroberts2002,goossens2002}. The present paper deals with this specific version of coronal seismology. Some recent reviews where this method and its applications are discussed are, e.g., \citet{goossens2008},\citet{ruderman2009}, \citet{demoortel2012}, and \citet{arregui2012}.

\citet{roberts1984} showed early examples of the determination of the magnetic field strength in coronal loops using some observed periods of coronal waves, presumably related to standing kink waves, along with the theoretical expression of the period in a thin magnetic tube \citep[see, e.g.,][]{edwin1983}. \citet{roberts1984}  imposed a large density contrast between the loop and its environment and took a particular value for the loop density. Subsequent works have refined the method outlined by \citet{roberts1984} and have used more accurate observations. For example, \citet{nakariakov2001}  used the period of a transverse loop oscillation event observed with TRACE to estimate the Alfv\'en velocity in the loop after imposing the value of the density contrast. Then, \citet{nakariakov2001} roughly estimated the loop density using the emission measure and were able to give an approximate value for the magnetic field strength. Concerning the use of the damping time for seismology, \citet{rudermanroberts2002} used the value of the ratio $\tau_{\rm D}/P$ of an event reported by \citet{nakariakov1999}, along with the theoretical expression of $\tau_{\rm D}/P$ in the thin tube thin boundary (TTTB) approximation \citep[see also][]{hollwegyang1988}, to estimate the loop transverse inhomogeneity lengthscale. A similar approach was followed by \citet{goossens2002}, who inferred  the transverse inhomogeneity lengthscale in a wider set of eleven oscillating coronal loops reported by \citet{ofman2002}. A limitation of the estimates computed by  \citet{rudermanroberts2002} and \citet{goossens2002}  is that  the value of the density contrast was arbitrarily imposed in both works. Subsequently, \citet{arregui2007seis} significantly improved these previous works by using observed values of both $P$ and $\tau_{\rm D}/P$ and by keeping the density contrast as an unknown quantity. They also abandoned the analytic TTTB approximation and used the fully numerical eigenvalue results of \citet{vandoorsselaere2004}. \citet{arregui2007seis} showed that the valid loop models inferred from the numerical seismic inversion describe a one-dimensional curve in the space of parameters formed by the Alfv\'en velocity, the transverse inhomogenity lengthscale, and the density contrast. Among these three parameters, the Alfv\'en velocity is the seismic variable that can be constrained the best, while the density contrast and the inhomogenity lengthscale remain unconstrained. Although \citet{arregui2007seis} showed that the Alfv\'en velocity can be constrained, no accurate determination of the magnetic field strength is possible unless a reliable value of the loop density is independently provided, and vice versa. Later, \citet{goossens2008seis} presented a simpler analytic inversion scheme  based on the TTTB approximation.  The analytic scheme of \citet{goossens2008seis} showed a good agreement with the numerical inversions of the events analyzed by \citet{arregui2007seis}. Recent efforts include the use of statistical methods. In this direction, the works by \citet{arregui2011baye} and \citet{asensio2013} used Bayesian analysis to set a framework to obtain all the possible information by combining theory, previous understanding, and available data. A different statistical approach  was followed by \citet{verwichte2013}, where the emphasis was put on obtaining  constrained intervals for the seismic variables.

With the exception of the article by \citet{arregui2007seis}, all the seismology papers cited in the above paragraph heavily rely on the expression of $\tau_{\rm D}/P$ derived in the TTTB approximation \citep[e.g.,][]{rudermanroberts2002,goossens2002}. However, some studies \citep[e.g.,][]{aschwanden2003} suggest that coronal loops are probably fully inhomogeneous in the transverse direction, a result that jeopardizes the applicability of the TTTB approximation.  \citet{vandoorsselaere2004} investigated the error in $\tau_{\rm D}/P$ due to the TTTB approximation when used beyond its range of applicability and concluded that it is 25\% at most. The result of \citet{vandoorsselaere2004} is often cited as a justification for the use of the TTTB formula when the requisites for the approximation are not satisfied, i.e., when the loop is largely nonuniform. Although the conclusion by \citet{vandoorsselaere2004} apparently resolved the problem of the applicability of the TTTB approximation, the purpose of the present paper is to  reanalyze the accuracy of the TTTB approximation and to investigate its impact on seismology estimates. Our reason for tackling this task is that \citet{vandoorsselaere2004} restricted themselves to a specific variation of density in the transverse direction, namely a sinusoidal profile, and they did not explore the influence of other profiles. As explained in \citet{partI}, hereafter Paper~I, the influence of the shape of density variation is important when the loop is largely nonuniform.  We are concerned about the effect of the transverse density profile on seismology estimates. 

Here, we continue the investigation started in Paper~I about the theoretical behavior of transverse waves in nonuniform flux tubes. In the previous paper, we studied the differences between ideal and resistive kink modes. We showed that there are some fundamental differences between the ideal and resistive eigenfunctions, but both ideal and resistive solutions provide  the same theoretical periods and damping rates. The differences in the eigenfunctions may be important for energy computations. For seismology, however, no matter whether  ideal or  resistive solutions are used: the information needed to perform  seismic inversions is the same in both cases. Hence, we can use the method developed in Paper~I to compute theoretical values of period and damping rate. Complicated resistive eigenvalue computations are not needed. The first goal in the present paper is to determine the error associated with the TTTB approximation when different density profiles are considered in the transverse direction. We aim to know whether the small error estimated by \citet{vandoorsselaere2004} applies also to other density profiles apart from the sinusoidal variation. Once this task is achieved, the second goal is to study the impact of the transverse density profile on seismically inferred physical parameters in coronal loops. To this end, we revisit the numerical \citep{arregui2007seis} and analytic \citep{goossens2008seis} inversion schemes and compare the effect of the transverse density profile in both cases.

This paper is organized as follows. Section~\ref{sec:model} contains the description of the coronal loop model and the TTTB approximations to the period and damping time of kink oscillations. Later, we investigate in Section~\ref{sec:thick} the error due to the TTTB approximation when applied beyond its range of validity. The error is computed for three paradigmatic transverse density profiles. Then, we explore in Section~\ref{sec:seis} the specific influence of the density profile on both analytic and numerical seismic inversions. Finally, Section~\ref{sec:conclusion} contains the discussion and conclusions of this work.

\section{MODEL AND TTTB APPROXIMATION}
\label{sec:model}

The  model to represent a transversely nonuniform coronal loop is the same as in Paper~I. The equilibrium  is a straight magnetic cylinder of radius $R$ and length $L$ embedded in a uniform and infinite plasma. We denote as $r$, $\varphi$, and $z$  the radial, azimuthal, and longitudinal coordinates, respectively. The magnetic field is  constant and parallel to the axis of the cylinder. The density, $\rho$, is  uniform in the azimuthal and longitudinal directions and nonuniform in the radial direction, namely 
\begin{equation}
 \rho(r) = \left\{
\begin{array}{lll}
\rhoi, & \textrm{if} & r \leq R - l/2, \\
\rhotr(r), & \textrm{if} & R -l/2 < r < R +  l/2,\\
\rhoe, & \textrm{if} & r \geq R+l/2,
\end{array}
\right.
\end{equation}
where $\rhoi$ and $\rhoe$ are internal and external constant densities, with $\rhoi > \rhoe$, and $\rhotr(r)$ is a nonuniform density profile that continuously connects the internal plasma to the external plasma in a transitional layer of thickness $l$.  The limits $l/R = 0$ and $l/R=2$ represent a magnetic tube with a piecewise constant density and a tube fully inhomogeneous in the radial direction, respectively. As in Paper~I, we consider three different spatial variations for the density in the nonuniform layer, namely a sinusoidal profile,
\begin{equation}
\rho_{\rm tr}(r) = \frac{\rhoi}{2} \left[\left( 1 + \frac{\rhoe}{\rhoi} \right) -  \left( 1 - \frac{\rhoe}{\rhoi} \right) \sin \left(\frac{\pi}{l}(r-R) \right) \right], \label{eq:sin}
\end{equation}
a linear profile,
\begin{equation}
\rho_{\rm tr}(r) = \rhoi - \frac{\rhoi - \rhoe}{l}\left( r - R + \frac{l}{2} \right),
\end{equation} 
and a parabolic profile,
\begin{equation}
\rho_{\rm tr}(r) = \rhoi - \frac{\rhoi-\rhoe}{l^2} \left( r - R + \frac{l}{2} \right)^2.
\end{equation}  
Unfortunately, present-day observations do not have sufficient resolution to determine the true shape of the transverse transitional layer in coronal loops. We are aware that the true profile is certainly none of the three simple profiles used here. Of course, there are infinite possible profiles that could be considered.  The purpose for choosing these three paradigmatic profiles is to point out the effects introduced by assuming different density variations in the transverse direction.

Linear ideal MHD waves superimposed on the equilibrium state are governed by Equations~(1) and (2) of Paper~I. Wave perturbations are assumed proportional to $\exp \left( i k_z z + i m \varphi - i \omega t \right)$, where $k_z$ and $m$ are the longitudinal and azimuthal wavenumbers, respectively, and $\omega$ is the frequency. We set $m=1$ corresponding to kink modes, i.e., modes that displace the axis of the flux tube and move it as a whole. These are the modes that have been related to the observations of transverse loop oscillations. Standing and propagating waves are equivalent from the mathematical point of view. Standing waves have $k_z$ fixed and $\omega$ determined by the dispersion relation. Conversely, propagating waves have $\omega$ fixed and $k_z$ determined by the dispersion relation. Here, we study the longitudinally fundamental standing kink mode. To represent the line-tying of the perturbations at the ends of the flux tube, i.e., at the photosphere, we set $k_z = \pi/L$. Due to resonant damping, the frequency is complex, namely $\omega = \omega_{\rm R} + i\omega_{\rm I}$, where $\omega_{\rm R}$ and $\omega_{\rm I}$ are the real and imaginary parts, respectively. The wave period, $P$, and  the exponential damping time, $\tau_{\rm D}$, are computed as
\begin{equation}
P = \frac{2\pi}{\omega_{\rm R}},  \qquad \tau_{\rm D} =  \frac{1}{|\omega_{\rm I}|}.
\end{equation} 

In the thin tube thin boundary (TTTB) approximation, i.e., for $L/R \gg 1$ and $l/R \ll 1$, the analytic approximate expressions to the kink mode $P$ and $\tau_{\rm D}/P$ are \citep[see, e.g.,][]{rudermanroberts2002,goossens2002,goossens2008seis,partI}
\begin{eqnarray}
P &=& \tau_{\rm A,i} \sqrt{\frac{2\left( \rhoi+\rhoe \right)}{\rhoi}}, \label{eq:p} \\
\frac{\tau_{\rm D}}{P} &=& F \frac{R}{l} \frac{\rhoi+\rhoe}{\rhoi-\rhoe}. \label{eq:tttb}
\end{eqnarray}
where $\tau_{\rm A,i} = L/\vai$ is the internal Alfv\'en travel time, with $\vai$ the internal Alfv\'en velocity, and $F$ is a numerical factor that depends of the specific density variation in the nonuniform layer. According to Equation~(\ref{eq:p}), $P$ is unaffected by the presence of the nonuniform layer. The factor $F$ in Equation~(\ref{eq:tttb}) is the only effect of the density profile that remains in the TTTB approximation to $\tau_{\rm D}/P$.  The formula to compute the factor $F$ is
\begin{equation}
F = \frac{4}{\pi^2} \frac{l}{\rhoi - \rhoe} \left| \frac{\der \rho}{ \der r} \right|_{\ra}, \label{eq:ffactor}
\end{equation}
where $\left|\der \rho / \der r\right|_{\ra}$ denotes the absolute value of the radial derivative of $\rho(r)$ evaluated at the Alfv\'en resonance position  $r=\ra$. The position of the resonance is determined by solving the equation
\begin{equation}
\rho_{\rm tr}(\ra) = \frac{\rhoi + \rhoe}{2}. \label{eq:resonance}
\end{equation}
A straightforward calculation shows that $\ra = R$ for both sinusoidal and linear profiles, and $\ra =R + \frac{\sqrt{2}-1}{2} l$ for the parabolic profile. Consequently, $F=2/\pi$ for the sinusoidal profile, $F=4/\pi^2$ for the linear profile, and $F=4\sqrt{2}/\pi^2$ for the parabolic profile. Note that in Paper~I we took $\ra = R$ in all cases to simplify matters, although that assumption may be inaccurate when the profile is asymmetric as in the parabolic case (J.~Andries, private communication). Here we consider a more appropriate value of the resonance position for the parabolic profile.

Results beyond the TTTB approximation obtained in Paper~I (see Figure~5 of Paper~I) showed that the shape of the nonuniform layer has an impact on both $P$ and $\tau_{\rm D}/P$ that is not predicted by the TTTB approximation. Specifically, $P$ is found to decrease ($\omega_{\rm R}$ increases) when $l/R$ increases. Nonuniformity affects the period so that the kink mode of a nonuniform loop has a shorter period than that of a uniform loop with the same length and density contrast. Also, the dependence of $\tau_{\rm D}/P$ on $l/R$ is more complicated than the simple dependence of Equation~(\ref{eq:tttb}). The effect of the transverse density profile for $l/R$  beyond the TTTB limit is not restricted to the factor $F$ present in Equation~(\ref{eq:tttb}). 

In this work we assume that the damping of transverse loop oscillations is caused by resonant absorption. Other possible mechanisms that may be involved in the damping  process are thought to be of minor importance and are ignored here. For instance, it has been shown that in thin magnetic tubes the damping of kink modes due to wave leakage is much less efficient than the damping due to resonant absorption \citep[see, e.g.,][]{spruit1982,goossens1993,vandoorsselaere2009}. Hence, the theoretical expression of the ratio $\tau_{\rm D}/P$ given in Equation~(\ref{eq:tttb}) captures the effect of resonant absorption only.

\section{ERROR ASSOCIATED WITH THE TTTB APPROXIMATION}
\label{sec:thick}

  \begin{figure*}
\centering
\includegraphics[width=.89\columnwidth]{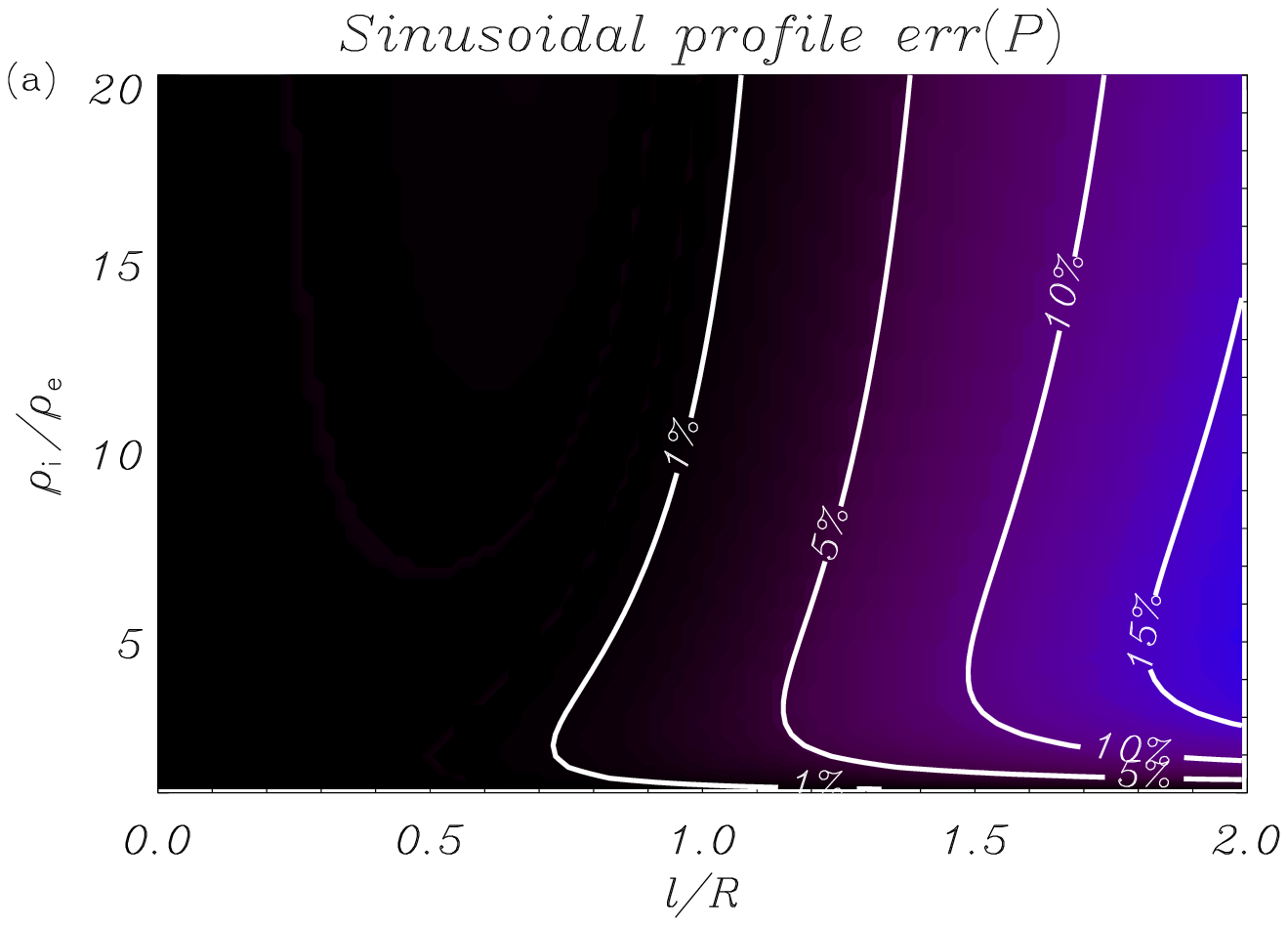}
\includegraphics[width=.89\columnwidth]{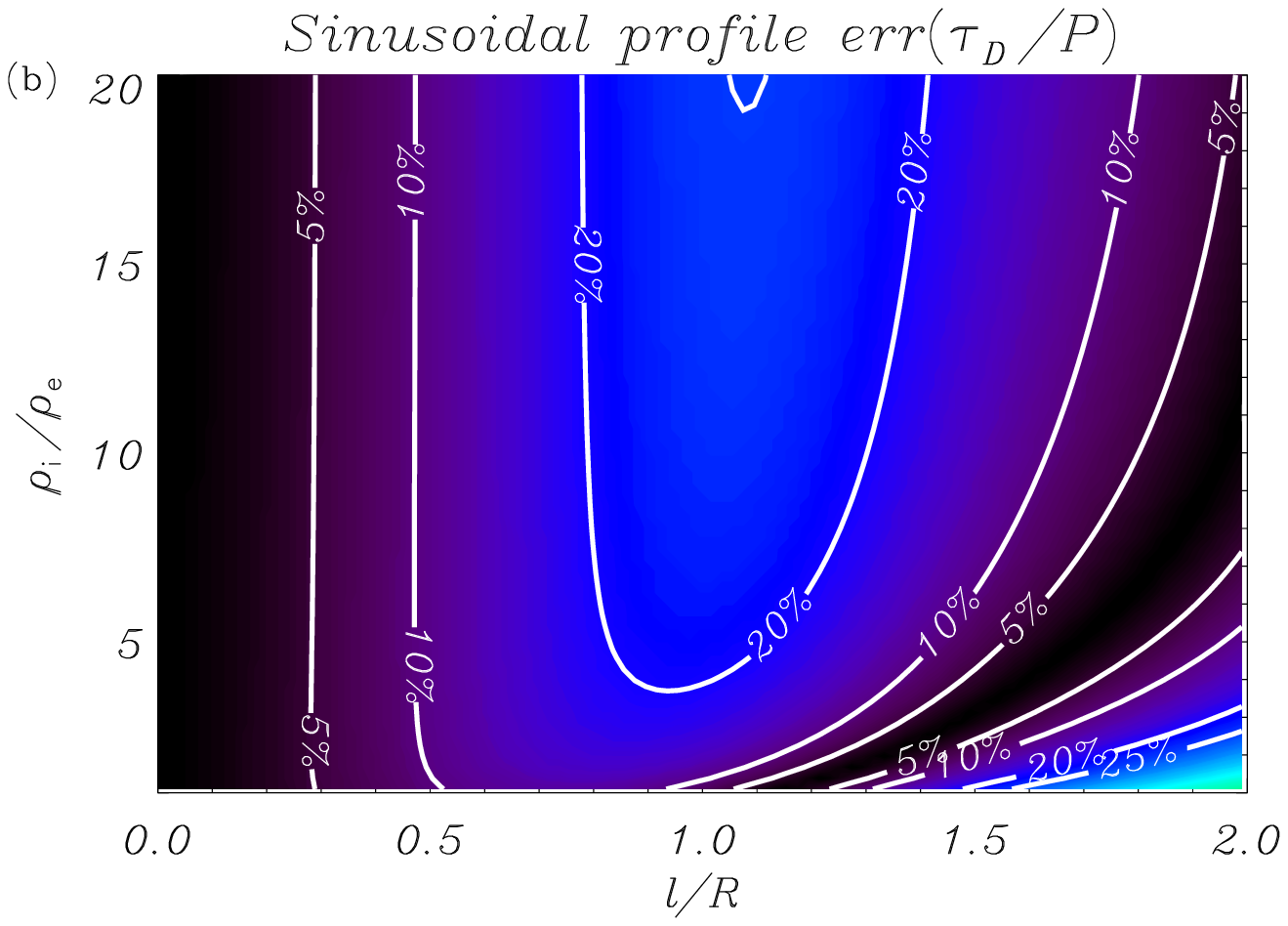}
	\includegraphics[width=.89\columnwidth]{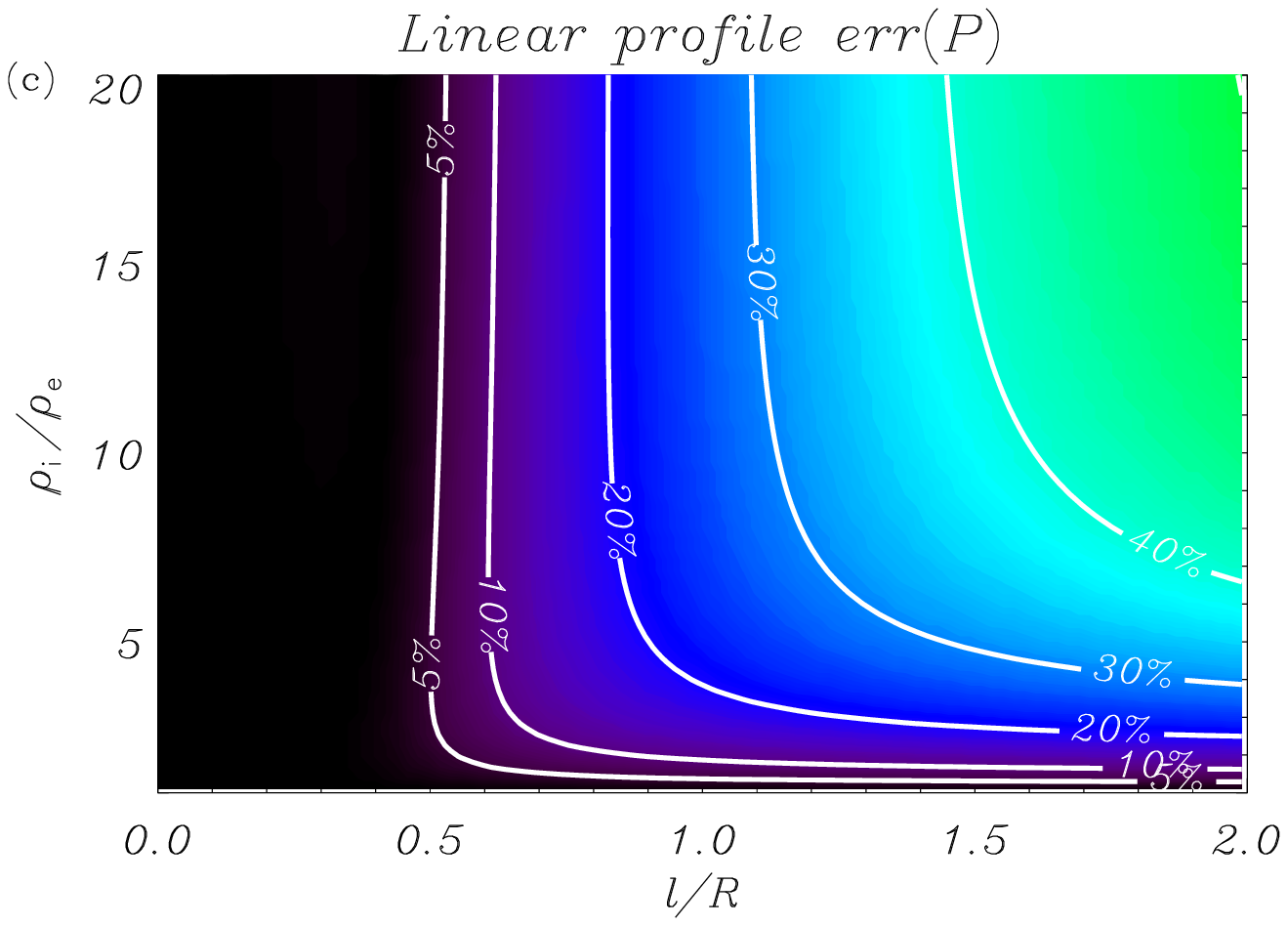}
			\includegraphics[width=.89\columnwidth]{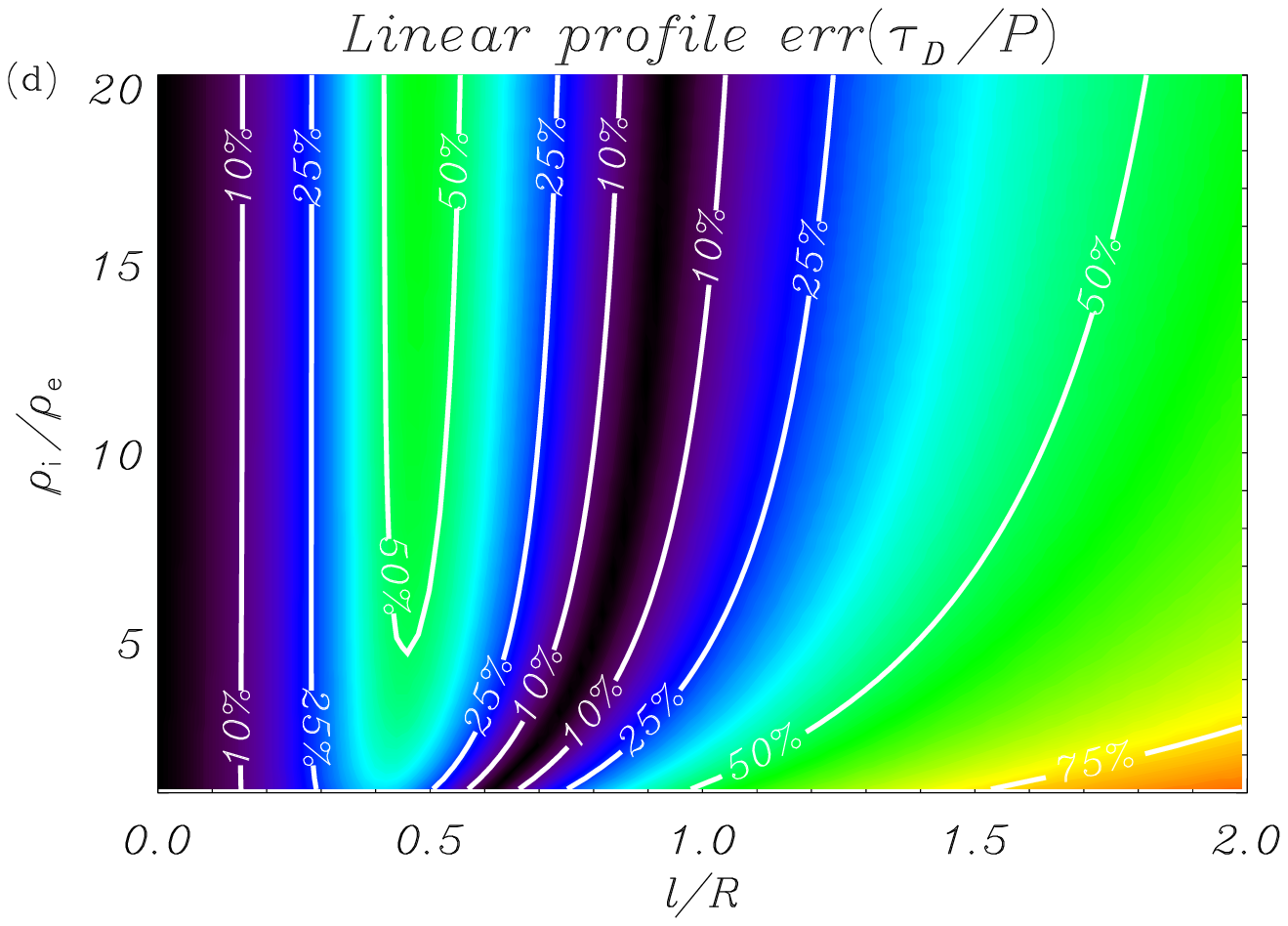}
			\includegraphics[width=.89\columnwidth]{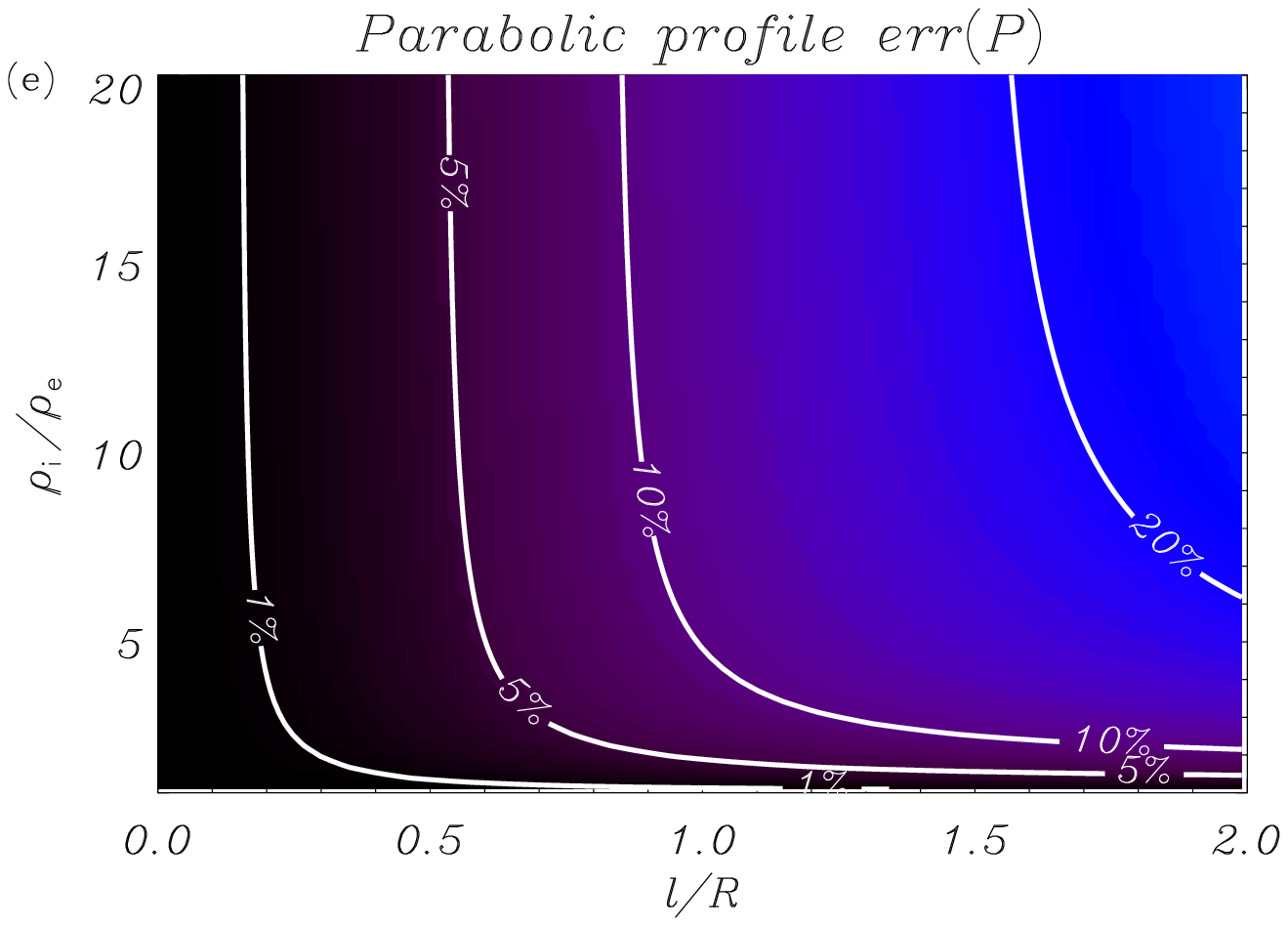}
			\includegraphics[width=.89\columnwidth]{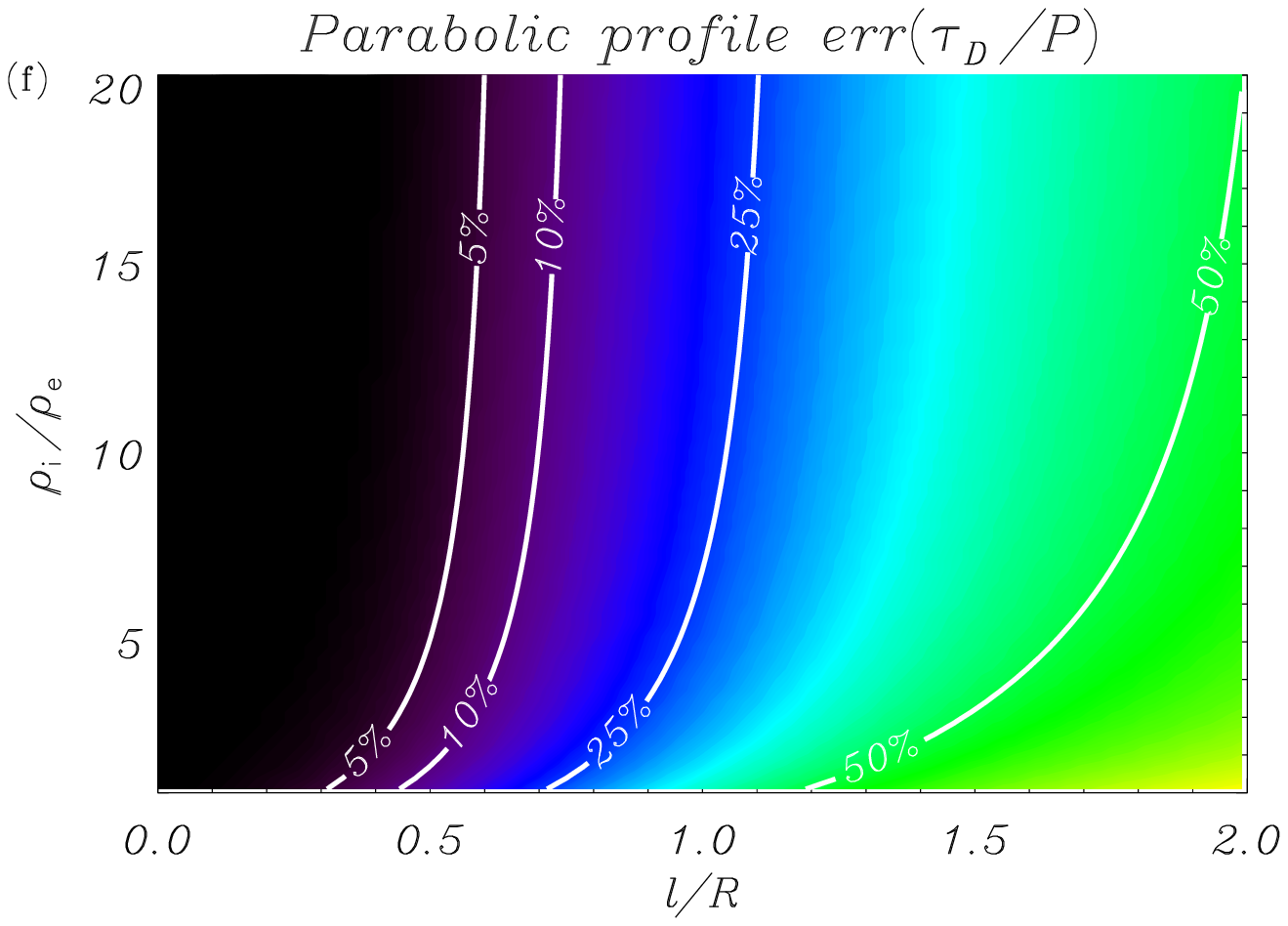}
			\includegraphics[width=1.8\columnwidth]{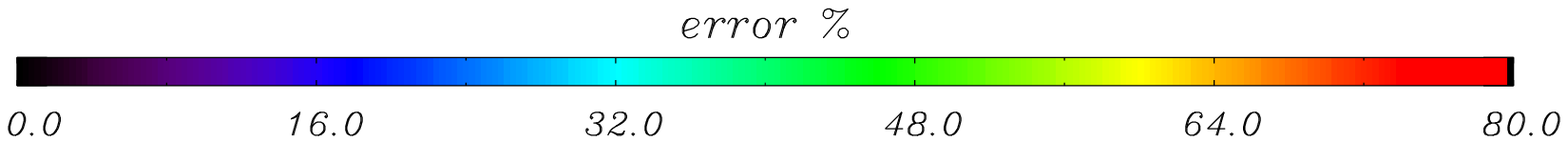}
	\caption{Contour plots  in the $l/R$-$\rhoi/\rhoe$ plane of parameters of the error of $P$ (left) and $\tau_{\rm D}/P$ (right) due to the use of the TTTB approximation for the sinusoidal (top), linear (mid), and parabolic (bottom) density profiles. In all cases we use $L/R = 100$. }
	\label{fig:sinerror}
\end{figure*}

In this Section, we put the TTTB approximation to the test when used beyond its theoretical range of validity. To this end, we compare the approximations to $P$ and $\tau_{\rm D}/P$ given in Equations~(\ref{eq:p}) and (\ref{eq:tttb}), respectively, to their actual values, which are computed by solving the  dispersion relation obtained in Paper~I (Equation~(27) of Paper~I). We refer the reader to Paper~I for extensive details about the derivation of the dispersion relation. The dispersion relation of Paper~I is valid for arbitrary values of $l/R$, while  Equations~(\ref{eq:p}) and (\ref{eq:tttb}) are  strictly valid when $l/R \ll 1$ only. 

For the following analysis, we define the error as
\begin{equation}
{\rm err}\left(X\right) \equiv \frac{|X-X_{\rm TTTB}|}{X},
\end{equation}
where $X$ denotes the exact $P$ or $\tau_{\rm D}/P$ computed from the  dispersion relation and $X_{\rm TTTB}$ denotes the corresponding TTTB approximations to these quantities (Equations~(\ref{eq:p}) and (\ref{eq:tttb})). We vary $\rhoi/\rhoe$ and $l/R$ in the ranges $\rhoi/\rhoe \in [1.1,20]$ and $l/R \in [0, 2]$ and compute the error of $P$ and $\tau_{\rm D}/P$. Figure~\ref{fig:sinerror} displays  contour plots of ${\rm err}(P)$ and ${\rm err}(\tau_{\rm D}/P)$ in the $(l/R,\rhoi/\rhoe)$-plane  for the three  density profiles used in this work. In the computations we use $L/R = 100$ based on observations that show that oscillating loops are roughly two orders of magnitude longer than their radii \citep[see, e.g.,][]{aschwanden2002}. \citet{vandoorsselaere2004} studied the dependence on $L/R$ (expressed as $k_z R$ in their article) and found a very weak dependence on this parameter as long as $L/R \gg 1$ (i.e., $k_z R \ll 1$). We have also considered smaller values of $L/R$ than that used in Figure~\ref{fig:sinerror} and no significant differences have been obtained. In the following paragraphs we discuss the results shown in Figure~\ref{fig:sinerror}.

We start by analyzing the error in $P$. The results for the three profiles have a similar dependence on $\rhoi/\rhoe$ and $l/R$. The error in $P$ shows a strong dependence on the density contrast when this parameter takes low values, i.e., for $\rhoi/\rhoe \lesssim 5$. For higher values of $\rhoi/\rhoe$, the density contrast weakly affects ${\rm err}(P)$, which is then mainly determined by $l/R$.   When thick transitional layers are considered, the linear profile produces the largest ${\rm err}(P)$, followed by the parabolic profile and, finally, the sinusoidal profile. For instance, when $\rhoi / \rhoe = 10$ and $l/R = 1$, ${\rm err}(P) \approx 20\%$ for the linear profile, ${\rm err}(P) \approx 10\%$ for the parabolic profile, and ${\rm err}(P) \approx 1\%$ for the sinusoidal profile. For a fully nonuniform tube, i.e., $l/R \approx 2$, ${\rm err}(P) \approx 45\%$ for the linear profile, ${\rm err}(P) \approx 23\%$ for the parabolic profile, and ${\rm err}(P) \approx 15\%$ for the sinusoidal profile using again  $\rhoi / \rhoe = 10$.  In all cases, the TTTB approximation to $P$ overestimates the actual value. Nonuniformity tends to produce shorter periods than those of uniform loops.

We turn to the error in $\tau_{\rm D}/P$. The three profiles produce significantly different results. We comment the differences between the results for the three profiles below:
\begin{enumerate}
 \item For the sinusoidal profile (Figure~\ref{fig:sinerror}(b)), we find that ${\rm err}(\tau_{\rm D}/P)$ increases with $l/R$ and is almost independent of the density contrast when $l/R \lesssim 1$. ${\rm err}(\tau_{\rm D}/P)$ reaches a maximum value of $25\%$, approximately, at $l/R \approx 1$. Later, ${\rm err}(\tau_{\rm D}/P)$ decreases again for $l/R\gtrsim 1$. Here, our results  are in good agreement with \citet{vandoorsselaere2004}, who also reported $25\%$ error in $\tau_{\rm D}/P$ for $l/R = 1$. The absolute maximum of the error is, approximately, $45\%$ and takes place near the bottom right corner of Figure~\ref{fig:sinerror}(b), where both $l/R$ and $\rhoi / \rhoe$ reach extreme values. It is relevant that ${\rm err}(\tau_{\rm D}/P)$ is not a monotonic function of $l/R$. Certain combinations of $l/R$ and $\rhoi/\rhoe$ cause ${\rm err}(\tau_{\rm D}/P)=0$ when $l/R > 1$. At these locations the actual damping rate coincides with TTTB approximate value. This was also noted by \citet{vandoorsselaere2004}.
 
 \item In the case of the linear profile (Figure~\ref{fig:sinerror}(d)),  ${\rm err}(\tau_{\rm D}/P)$ shows again a complicated dependence with the thickness of the nonuniform layer. ${\rm err}(\tau_{\rm D}/P)$ increases when $l/R$ increases from zero until ${\rm err}(\tau_{\rm D}/P) \approx 50\%$ is reached when $l/R \approx 0.5$. Then, ${\rm err}(\tau_{\rm D}/P)$ decreases again and becomes negligible when $l/R \approx 0.8$--$0.9$. After this interval, ${\rm err}(\tau_{\rm D}/P)$ increases again with $l/R$. For $l/R > 1.5$, ${\rm err}(\tau_{\rm D}/P)$ is  larger than $50\%$.

 \item The error of $\tau_{\rm D}/P$ for the parabolic profile (Figure~\ref{fig:sinerror}(f)) displays a simpler dependence with $l/R$ than in the other two profiles. The parabolic profile is the profile for which the TTTB approximation to $\tau_{\rm D}/P$ works the best for relatively thin layers. For example, when $l/R\approx 0.5$ the error of $\tau_{\rm D}/P$ is only about 5\%. The error keeps increasing monotonically as $l/R$ increases. For very thick layers the error of $\tau_{\rm D}/P$ for the parabolic profile is larger than that for the sinusoidal profile but smaller than that for the linear profile.

\end{enumerate}

In summary, we find that the specific form of the density variation across the loop influences the error associated with the TTTB approximation to both $P$ and $\tau_{\rm D}/P$, although the largest errors are found in the case of $\tau_{\rm D}/P$. Among the three paradigmatic profiles considered here, the  error due to the use of the TTTB approximation is smallest for the sinusoidal variation when thick transitional layers are considered. Conversely, the parabolic profile produces smaller errors for relatively thin layers than any of the other two profiles. Importantly, the maximal $25\%$ error in $\tau_{\rm D}/P$  estimated by \citet{vandoorsselaere2004} only applies to a sinusoidal variation of density. For other density profiles, the maximal error due to the use of the TTTB approximation  beyond its range of applicability can be much larger. It is also possible that other density profiles different from the ones used here may produce maximal errors smaller than $25\%$.  The accuracy of the TTTB approximation is found to be strongly related to the specific density variation considered.

The results above indicate that the  shape of the nonuniform layer in coronal loops, and therefore our ignorance of this parameter, is very relevant for the applicability of the TTTB approximation. The relatively small error estimated by  \citet{vandoorsselaere2004}, which was computed in that paper for  a specific density variation, should not be used as a general result to justify the use of the TTTB approximation beyond its range of applicability in all cases. 

\section{IMPACT ON SEISMOLOGY ESTIMATES}
\label{sec:seis}

\begin{figure*}
		\centerline{\includegraphics[width=.85\columnwidth]{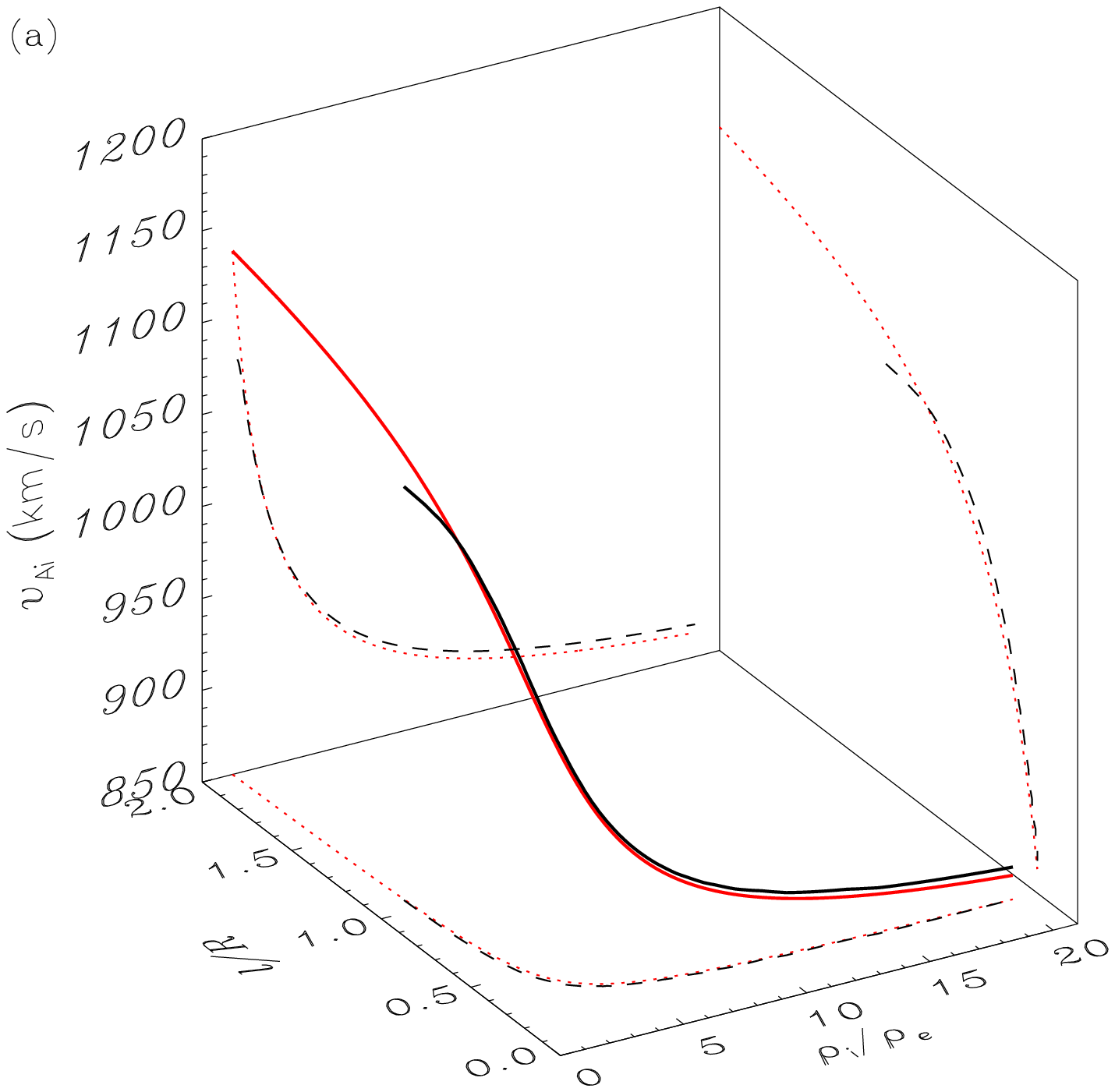}
		\includegraphics[width=.85\columnwidth]{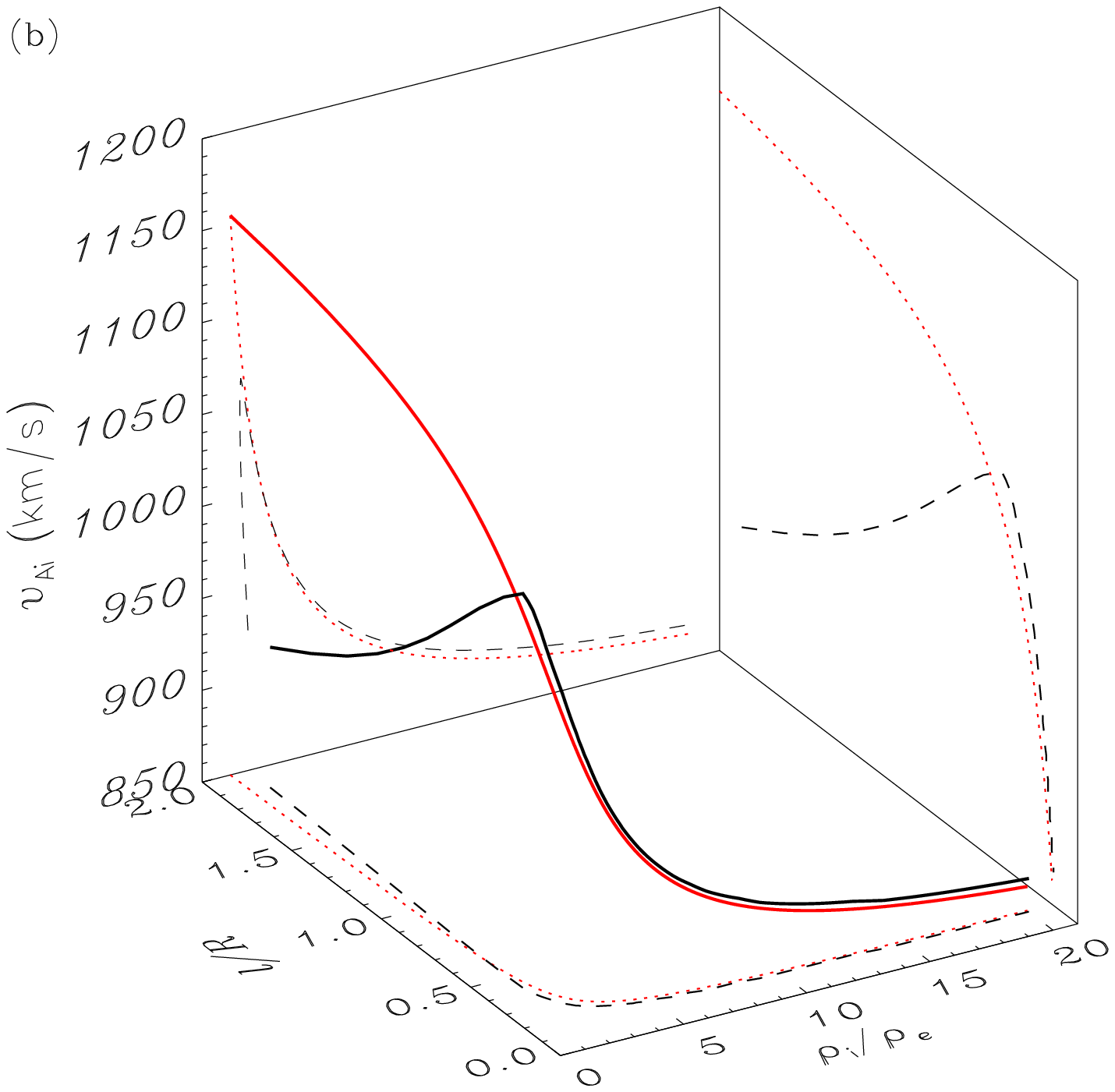}}
		\centerline{\includegraphics[width=.85\columnwidth]{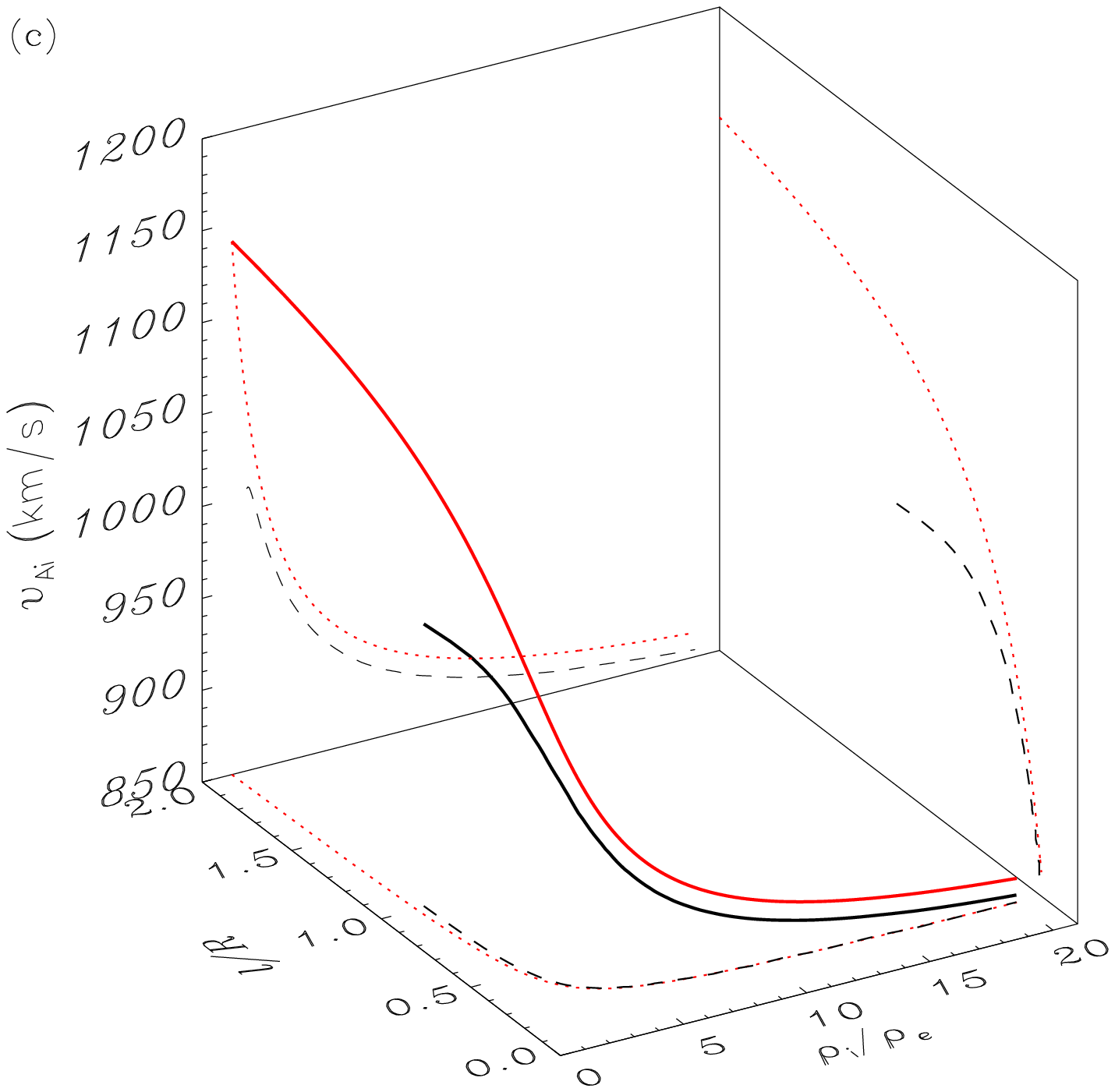}
		\includegraphics[width=.85\columnwidth]{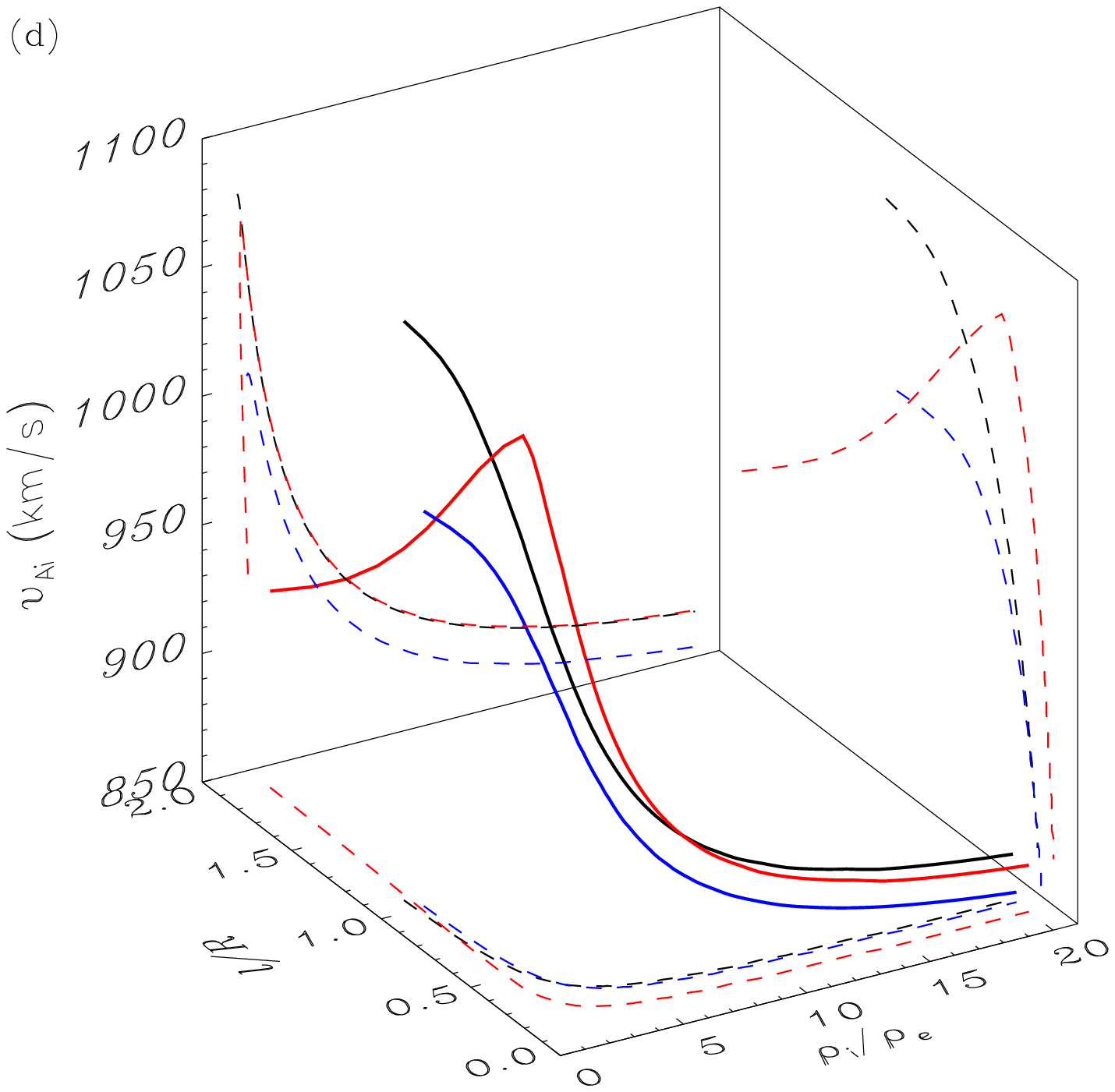}}
	\caption{Result of the seismological inversion of event \#5 of \citet{ofman2002} for (a) the sinusoidal density profile, (b) the linear density profile, and (c) the parabolic density profile. The  black solid lines are the actual numerical inversions while the red solid lines are the approximate analytic  inversions. The discontinuous lines are the projections of the solutions to the various planes. Panel (d) shows a comparison between the  numerical inversions for the sinusoidal (black), linear (red), and parabolic (blue) profiles.}
	\label{fig:inversion}
\end{figure*}

\begin{figure*}
		\centerline{\includegraphics[width=.85\columnwidth]{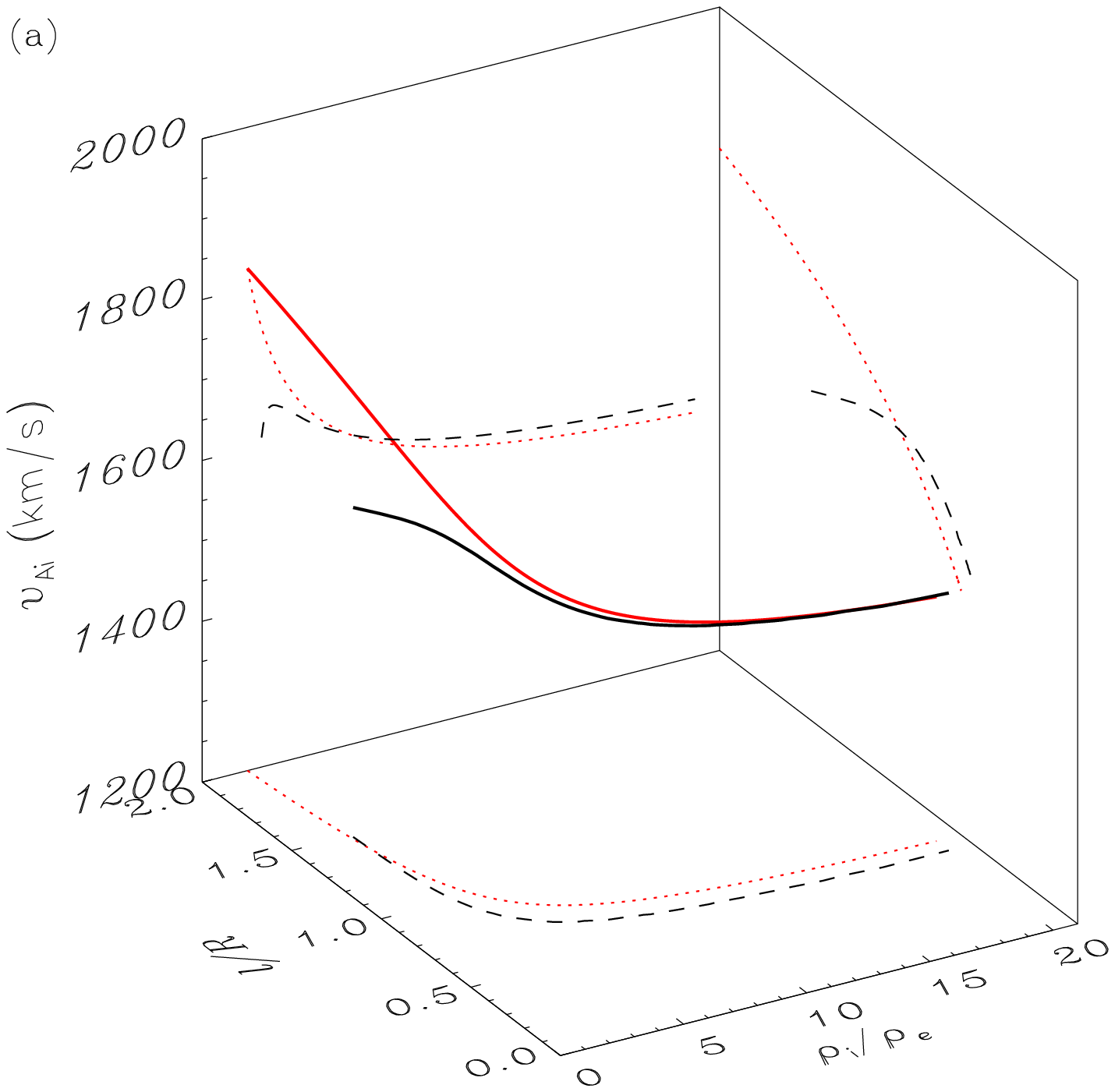}
		\includegraphics[width=.85\columnwidth]{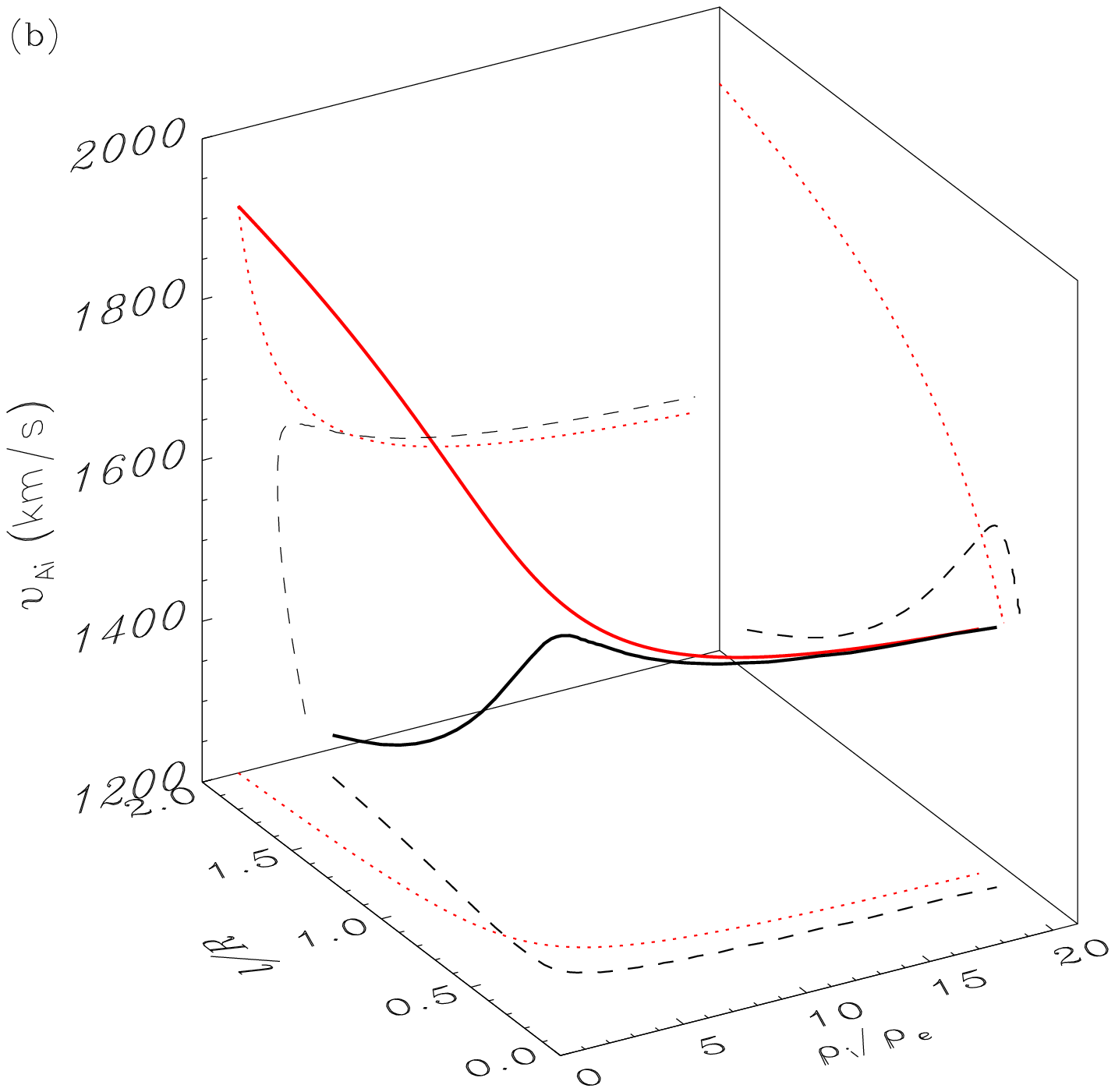}}
		\centerline{\includegraphics[width=.85\columnwidth]{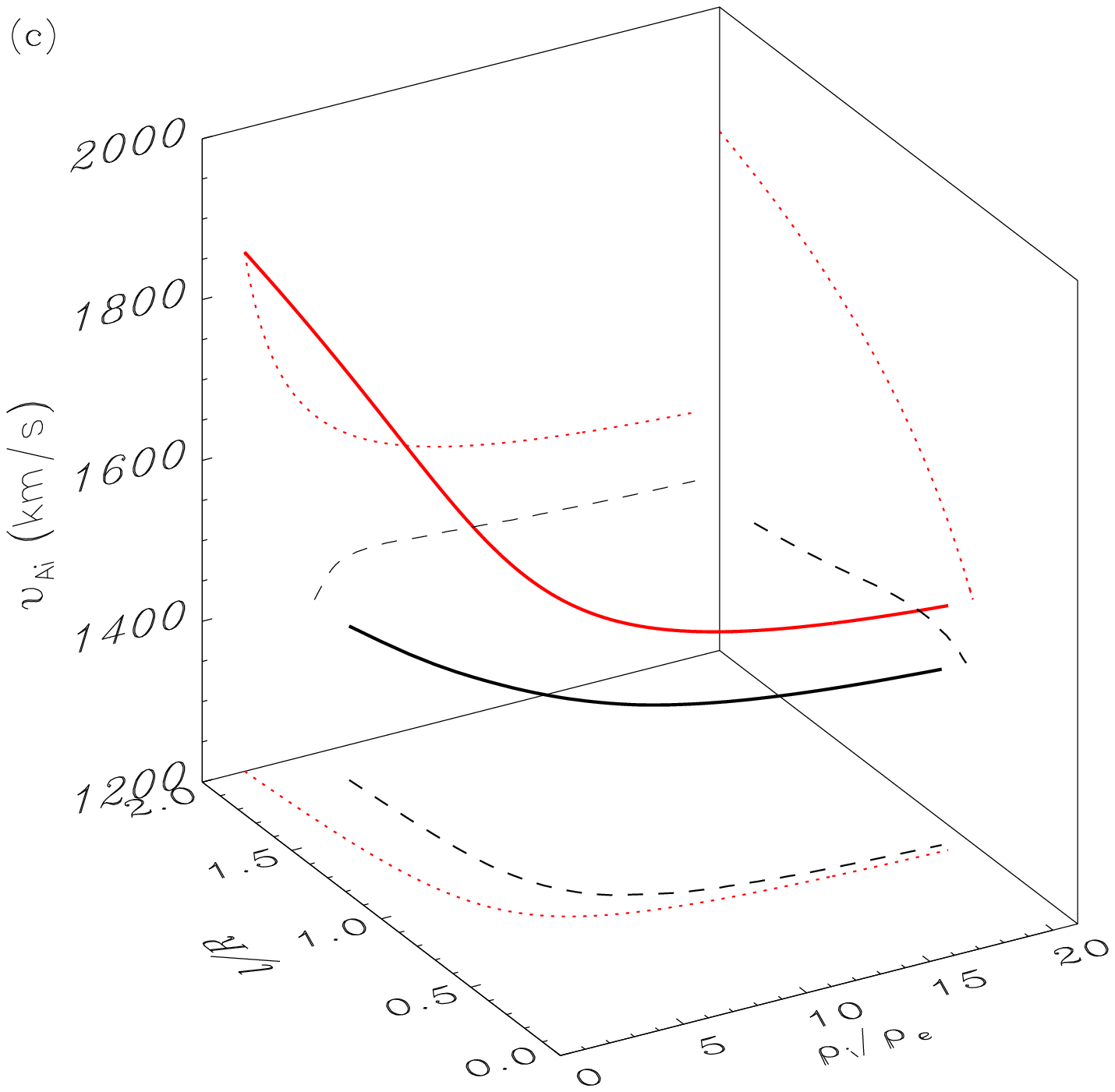}
		\includegraphics[width=.85\columnwidth]{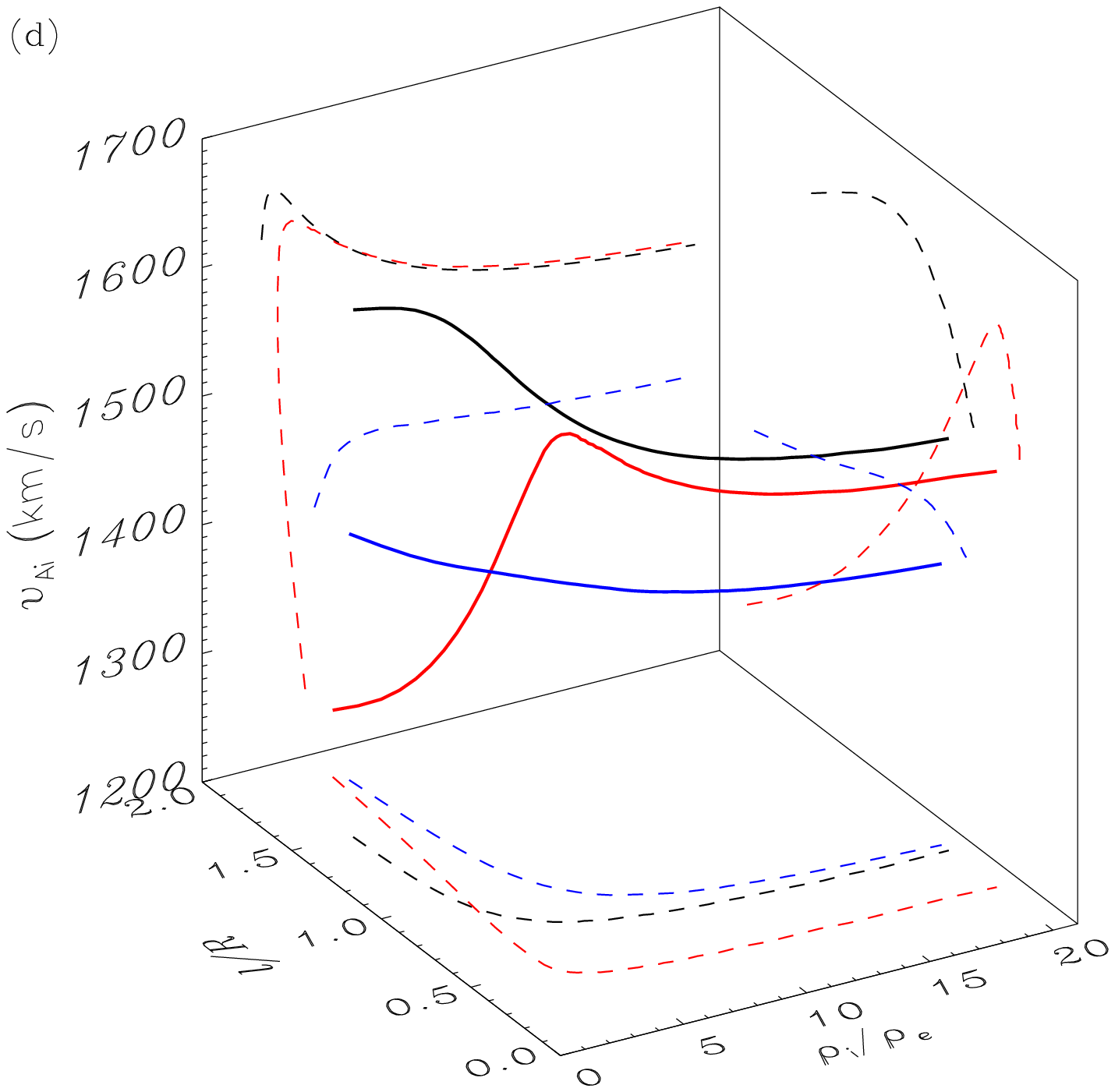}}
	\caption{Same as Figure~\ref{fig:inversion} but for event \#10 of \citet{ofman2002}.}
	\label{fig:inversion2}
\end{figure*}

Here we explore how the error associated with the TTTB approximation affects the inversion of physical parameters in coronal loops using inversion schemes based on this approximation. In addition, we investigate the influence of the shape of the nonuniform layer on the inversions. 
   
First, we briefly introduce the analytic  inversion scheme based on the TTTB approximation. This scheme was presented by \citet{goossens2008seis} for standing waves and by  \citet{goossens2012seis} for propagating waves. Here we use the scheme for standing waves.  By combining  Equations~(\ref{eq:p}) and (\ref{eq:tttb}), it is possible to express the internal Alfv\'en velocity, $\vai$, and the inhomogeneity lengthscale, $l/R$, as functions of the density contrast, $\rhoi/\rhoe$, as follows \citep{goossens2008seis}
\begin{eqnarray}
\vai &=& \frac{L}{P} \sqrt{\frac{2\left(1+\rhoi/\rhoe\right)}{\rhoi/\rhoe}}, \label{eq:vai} \\
\frac{l}{R} &=&  \frac{1}{C} \frac{\rhoi/\rhoe+1}{\rhoi/\rhoe-1}, \label{eq:lor}
\end{eqnarray}
where the parameter $C$ is defined as
\begin{equation}
C = \frac{1}{F}\frac{\tau_{\rm D}}{P}.
\end{equation}  
\citet{goossens2008seis} defined $C$ for the specific case of a sinusoidal transition in density. Here we have generalized the definition  by including the factor $F$, which allows us to consider any density profile. In turn, from Equation~(\ref{eq:lor}) we get that $\rhoi/\rhoe$ is related to $l/R$ by
\begin{equation}
\frac{\rhoi}{\rhoe} = \frac{ \left(l/R\right) C + 1}{\left(l/R\right) C -1}.
\end{equation}

The quantities that are assumed to be known from the observations are $P$, $\tau_{\rm D}$, and $L$.  Conversely, $\vai$, $l/R$, and $\rhoi/\rhoe$ are unknown and are the  variables to be inferred from the seismological inversion. \citet{goossens2008seis} assumed that $L$ was also unknown, so they used the internal Alfv\'en travel time $\tau_{\rm A,i} = L/\vai$ instead of $\vai$ as a seismic variable. Since the length of the loop is a quantity that is often reported by the observations, here we assume that $L$ is known and use $\vai$ as a seismic variable. We have only two equations (Equations~(\ref{eq:vai}) and (\ref{eq:lor})) that relate the three unknown quantities $\vai$, $l/R$, and $\rhoi/\rhoe$ to the three observed quantities $P$, $\tau_{\rm D}$, and $L$. As a consequence, there are infinite solutions. The whole collection of values of the seismic variables that are compatible with the observations describe a one-dimensional curve in the three-dimensional space of variables. The seismic variables are constrained in the following intervals,  
\begin{eqnarray}
\vai & \in & \left] \sqrt{2} \frac{L}{P}, \frac{2 L}{P} \right], \label{eq:intva}\\
\frac{l}{R} & \in & \left] \frac{1}{C} , 2 \right], \label{eq:intlor}\\
\frac{\rhoi}{\rhoe} & \in & \left[  \frac{2 C + 1}{2 C - 1} , \infty \right[. \label{eq:intcon}
\end{eqnarray}  
 A priori, with no additional information about the uncertainties of the observed quantities, any point on the  one-dimensional inversion curve is equally compatible with the observations \citep{arregui2007seis,goossens2008seis}.  
 
 We stress that the density profile is in principle unknown, so that the value of $F$ has to be assumed ad hoc. This is an important fact whose implications have been usually ignored in previous works. If $F$ were kept undetermined, the solution of the seismic problem would be even more complex since $F$ should be considered as an additional unknown. Then, the seismic variables would span a four-dimensional space.

In addition to the analytic inversion, we also perform the full numerical inversion \citep[see][]{arregui2007seis}. This is done by using the numerical solutions of the general dispersion relation (Equation~(27) of Paper~I) in the whole space of parameters. A difference with the analytic approximate inversion is that the radius of the loop, $R$, is also needed in the numerical inversion. With the values of $L$ and $R$ provided by the observations, we solve the dispersion relation by varying the seismic parameters $l/R$ and $\rhoi/\rhoe$ in the ranges $l/R \in \left[ 10^{-3}, 2  \right]$ and $\rhoi / \rhoe \in \left[ 1.1, 20 \right]$. Then, the procedure to perform the numerical inversion is similar to the one followed by \citet{arregui2007seis}. The only difference is that  \citet{arregui2007seis} used the numerical solutions of the resistive eigenvalue problem computed by \citet{vandoorsselaere2004}, and here we use the solutions of the ideal dispersion relation. As shown in Paper~I, both methods provide the same values of period and damping time of kink modes. The procedure is as follows.  We compute the theoretical $\tau_{\rm D}/P$ for each couple $(l/R, \rhoi/\rhoe)$ and determine the collection of couples $(l/R, \rhoi/\rhoe)$ whose theoretical $\tau_{\rm D}/P$ coincides with the observed value. This collection of couples forms a curve in the $(l/R, \rhoi/\rhoe)$-plane \citep[see, e.g.,][Figure~2]{arregui2007seis}. For each point on this curve, we require the  theoretical $P$ to be equal to the observed $P$. This determines the value of $\vai$ corresponding to each solution couple $(l/R, \rhoi/\rhoe)$, so that a one-dimensional inversion curve in the three-dimensional  $(\vai,l/R, \rhoi/\rhoe)$-space is formed.

The analytic and numerical inversions are performed for two of the eleven transverse loop oscillation events reported by \citet{ofman2002}, namely events \#5 and \#10. We choose these two events for two reasons. (1) They were already analyzed in previous papers \citep[e.g.,][]{arregui2007seis,goossens2008seis,arregui2011baye} for a sinusoidal transition in density. (2) They correspond to observations of moderate (event \#5, $\tau_{\rm D}/P \approx 3.12$) and strong (event \#10, $\tau_{\rm D}/P \approx 1.08$) damping, so that we can compare the analytic and numerical inversions in these two situations. The relevant parameters needed for the inversions are given in Table~1 of \citet{ofman2002}, which we reproduce here:  $L = 1.62\times 10^8$~m, $R=3.65\times 10^6$~m, $P=272$~s, and $\tau_{\rm D} = 849$~s for event \#5, and  $L = 1.92\times 10^8$~m, $R=3.45\times 10^6$~m, $P=185$~s, and $\tau_{\rm D} = 200$~s for event \#10.  The result of the inversions of events \#5 and \#10 are displayed in Figures~\ref{fig:inversion} and \ref{fig:inversion2}, respectively. These results are discussed in the following subsections.

\begin{deluxetable*}{cccccccc}
\tabletypesize{\scriptsize}

\centering

\tablecaption{Intervals of the seismic variables $\vai$, $l/R$, and $\rhoi/\rhoe$ for events \#5 and \#10 of \citet{ofman2002} obtained from the analytic  inversion (TTTB) and the numerical inversion (N) and for the three paradigmatic density profiles considered. \label{tab:summ}}

\tablenum{1}

\tablehead{\colhead{Event} &\colhead{Profile} & \colhead{$v_{\rm A,i,TTTB}$} & \colhead{$v_{\rm A,i,N}$} & \colhead{$\left(l/R\right)_{\rm TTTB}$} & \colhead{$\left(l/R\right)_{\rm N}$} & \colhead{$\left(\rhoi/\rhoe \right)_{\rm TTTB}$} & \colhead{$\left(\rhoi/\rhoe \right)_{\rm N}$} } 

%% All data must appear between the \startdata and \enddata commands
\startdata 
\#5 & Sinusoidal  &  842--1191 & 868--1075 & 0.20--2 & 0.22--1.07 & 1.23--$\infty$ & 1.43--20 \\
\#5 & Linear  &  842--1191 & 868--1064 & 0.13--2 & 0.14--1.88 & 1.14--$\infty$ & 1.53--20 \\
\#5 & Parabolic  &  842--1191 & 854--1004 & 0.18--2 & 0.20--1.01 & 1.20--$\infty$ & 1.80--20 \\
\hline
\#10 & Sinusoidal  &  1468--2076 & 1520--1646 & 0.59--2 & 0.58--1.49 & 1.84--$\infty$ & 2.41--20 \\
\#10 & Linear  &  1468--2076 & 1252--1619 & 0.38--2 & 0.31--1.85 & 1.46--$\infty$ & 3.06--20 \\
\#10 & Parabolic  &  1468--2076 & 1391--1439 & 0.53--2 & 0.62--1.81 & 1.72--$\infty$ & 4.56--20
\enddata

\tablecomments{$\vai$ is given in km~s$^{-1}$. The upper bound of $\left(\rhoi/\rhoe \right)_{\rm N}$ is taken to be the maximum value of $\rhoi/\rhoe$ used in the numerical inversions.}

\end{deluxetable*}
  
\subsection{Comparison of analytic and numerical inversion curves}  
  
First, we visually compare the analytic and numerical inversion curves. For event \#5 (Figure~\ref{fig:inversion}(a)--(c)), we find a reasonably good agreement between analytic and numerical inversions except in the case of the linear profile when low density contrast and thick layers are considered. For event \#10 (Figure~\ref{fig:inversion2}(a)--(c)), the agreement between the analytic and numerical curves is poorer than for event \#5. In both events, the sinusoidal profile provides the best agreement.

As pointed out by \citet{goossens2008seis}, the analytic inversion curve is always monotonic, while the numerical curves may have a nonmonotonic behavior  \citep[see also Figure~3(b) of][]{arregui2007seis}. As a consequence, the analytic inversion is most accurate  when the numerical inversion curve is monotonic too.  Monotonic inversion curves are typically obtained in events with weak or moderate damping as, e.g., event \#5, although not for all kind of density profiles. For event \#5, the inversion curves in the cases of sinusoidal (Figure~\ref{fig:inversion}(a)) and parabolic (Figure~\ref{fig:inversion}(c)) profiles are monotonic, while the inversion curve in the case of the linear profile (Figure~\ref{fig:inversion}(b)) is nonmonotonic for low density contrast and thick layers. Conversely, we find that all inversion curves are nonmonotonic in strongly damped events with $\tau_{\rm D}/P \sim 1$, as event \#10, so that the analytic inversions are not so accurate for event \#10 as they are for event \#5. The reason is that strongly damped oscillations require relatively large values of $l/R$. Hence, the applicability of the TTTB approximation for strongly damped events is compromised.

\subsection{Valid intervals of the seismic variables}  

We  recall that, without additional information, any point on the inversion curve is equally compatible with the observations. Thus, the actual shape of the inversion curve is not so important as long as the valid intervals of the seismic variables remain unaltered. The useful seismic information is given by these valid intervals, which indicate  the degree to which the seismic variables can be constrained. We compare in Table~\ref{tab:summ} the valid intervals of the seismic variables obtained with the three density profiles for the two events studied here. These results are discussed below.

 In the case of the analytic inversions, the same conclusions  apply to both events \#5 and \#10. Table~\ref{tab:summ} reveals that, in the same event, the analytic TTTB intervals of the seismic variables for the three profiles are very similar. As shown in Equations~(\ref{eq:intva})--(\ref{eq:intcon}), in the analytic scheme the density profile can only possibly affect the lower boundaries of $l/R$ and $\rhoi/\rhoe$, although we find that there are no significant differences between the three profiles. In all cases, $l/R$ and $\rhoi/\rhoe$ remain poorly constrained in the analytic inversions.  The internal Alfv\'en velocity, $\vai$, is the variable that is constrained the best \citep{arregui2007seis}, but  the density profile has no impact at all on the constraint of $\vai$ using the analytic scheme. Hence, the specific form of the density profile has a negligible impact on the seismic intervals obtained with the analytic inversion.

The situation changes dramatically when we move to  actual numerical inversions. Figures~\ref{fig:inversion}(d) and \ref{fig:inversion2}(d) compare the various numerical inversion curves of events \#5 and \#10, respectively. First of all, we see in Table~\ref{tab:summ} that the upper boundary of $l/R$ is smaller than $l/R=2$ in the numerical inversions. This  causes the interval of $l/R$ to be more constrained in the numerical inversions than in the analytic inversions.  Also, the lower boundary of $\rhoi/\rhoe$ is larger in the numerical inversions than in the analytic inversions. Concerning the valid intervals of $\vai$, we find significant differences between the results for events \#5 and \#10. For the moderately damped event \#5, the valid intervals of $\vai$ are similar for the three profiles and reasonably agree to those of the analytic inversions. On the contrary, for the strongly damped event \#10, $\vai$  displays different intervals when moving from one profile to another, which points out that the density profile has a strong impact on the constraint of $\vai$ in the numerical inversions of event \#10. The sinusoidal and parabolic profiles produce quite narrow intervals of $\vai$, but these intervals do not overlap, i.e., the constraints of $\vai$ are mutually exclusive for these two profiles. Conversely, the linear profile produces a wide interval of $\vai$ that overlaps with the intervals of both sinusoidal and parabolic profiles. In all cases, the intervals of $\vai$ in the numerical inversions do not match those of the analytic inversions.
  
In summary, in this Section we find that the analytic TTTB scheme fails to take into account the full impact of the transverse density profile on the valid intervals of the seismic variables. Since this impact is not very important for weakly or moderately damped oscillations, as event \#5, the analytic intervals are quite accurate in that case. However, for strongly damped oscillations, as event \#10, the density profile plays an important role in constraining $\vai$  that is not captured by the analytic scheme.  It is also relevant that the seismic variables in the various numerical inversions of the same event are constrained in different intervals depending on the density profile assumed. In turn, the numerically obtained intervals are different from the intervals predicted by the analytic scheme.

\section{DISCUSSION}  

\label{sec:conclusion}  

In this paper we have continued the study initiated in Paper~I \citep{partI} about the behavior of kink MHD waves in transversely nonuniform solar flux tubes. In connection to coronal loop transverse oscillations, here we investigated the accuracy of the TTTB approximations to the period and damping rate of kink modes. The accuracy of the TTTB approximation was first investigated by  \citet{vandoorsselaere2004} for the specific case of a sinusoidal variation of density in the transverse direction. We revisited the case studied by \citet{vandoorsselaere2004} and  determined the influence of other density variations. In addition, we explored how the error associated with the TTTB approximation impacts on seismology inversion schemes.

We find that the accuracy of the TTTB approximation to $P$ and $\tau_{\rm D}/P$ is very sensitive to the density profile considered in the inhomogeneous layer.  Nonuniformity affects the period so that nonuniform loops have shorter periods than uniform loops. After studying three paradigmatic density profiles, we conclude that the error associated with the TTTB formula of $\tau_{\rm D}/P$ is typically larger than the 25\% error estimated by \citet{vandoorsselaere2004}.  Accidentally, the sinusoidal profile used by  \citet{vandoorsselaere2004} seems to minimize the error due to the use of the TTTB approximation when thick layers are considered. Depending on the density profile used, the deviation of the  approximate $\tau_{\rm D}/P$ from its actual value can be larger than 50\% even for relatively thin nonuniform layers. For fully nonuniform loops, the error in  $\tau_{\rm D}/P$ can be much larger. Unfortunately, the actual errors associated with the TTTB approximate formulas to $P$ and $\tau_{\rm D}/P$ are  uncertain because the radial structuring and  true shape of the nonuniform layer in  coronal loops is presently unknown due to the resolution limitations of current observations. 

The error due to the use of the TTTB approximation directly impacts on the accuracy of the analytic seismic inversions. On the one hand, the approximate inversions do not recover  the nonmonotonic behavior of the actual numerical inversion curves. On the other hand, the approximate inversions fail to take into account the specific influence of the transverse density profile on the valid intervals of the seismic variables. This last result is not very relevant in the case of weakly or moderately damped oscillations, because these events favor small values of $l/R$ for which the TTTB approximation is good. However, strongly damped events require values of $l/R$ departing from the range of validity of the TTTB approximation. In those cases, the intervals of the seismic variables show a strong dependence on the density profile, which is not recovered by the analytic inversion scheme.

The results summarized in the above paragraphs point out that the approximate TTTB inversion scheme should not be used in the case of strongly damped oscillations. In those events, the full numerical inversion is required. However, caution is needed even when using the numerical scheme. The main reason for caution is that an ad hoc density profile has to be assumed to perform the inversion, because  the actual transverse structuring in coronal loops is ignored. Thus, the inferred intervals of the seismic variables are directly affected by the specific choice of  density variation. This fact may dispute the reliability of the seismic intervals.  Note that if one assumes a density profile in the nonuniform layer, then the only well-constrained seismic variable is the internal Alfv\'en velocity. But our ignorance of the density variation between the loop and its environment leads to a poorly limited $\vai$ for strongly damped transverse oscillations. In those events, no seismic variable  can be reliably constrained if the density profile in the nonuniform layer is unknown. This uncertainty should be taken into account when  the results of seismic inversions are used to probe the coronal plasma.

Here, we have focused on seismology with standing waves, but our conclusions can be easily extended to propagating waves. \citet{goossens2012seis} generalized  the analytic seismology scheme of \citet{goossens2008seis} to the case of propagating waves. For propagating waves, the observables are the wavelength, the damping length, and the phase velocity, while the seismic variables are the same as for standing waves, namely $\vai$, $l/R$, and $\rhoi/\rhoe$. The seismic variables are related to the observables by equivalent expressions to those of standing waves. Thus, the impact of the transverse density profile on the determination of the seismic variables using propagating waves is the same as that using standing waves.

Recently, several works have used the analytic TTTB scheme along with   statistical techniques that combine data from many different observations. Although those works made extensive use of the TTTB approximation, the inferred values of $l/R$  do not satisfy the preliminary hypothesis that $l/R \ll 1$. The results of the present paper may put into question the accuracy of the seismic intervals inferred in those studies.   For example, \citet{verwichte2013} used 52 events of standing transverse loop oscillations and obtained that the maximum value of $l/R$ compatible with the observations is in the range $[0.98,2]$, with the best fit being $l/R = 1.2$. \citet{verwichte2013} also used observations by the Coronal Multichannel Polarimeter to analyze propagating transverse waves and concluded that the maximum value of $l/R$ in coronal waveguides is about 2.06 times its minimum value. They argued that it is expected from physical reasons and modelling that the waveguides are largely inhomogeneous, hence the value of $l/R$ would be in the range $l/R \in [0.97,2]$. Independently, \citet{asensio2013} used  a different approach based on a Bayesian hierarchical method to analyze 30 events of transverse loop oscillations reported by \citet{aschwanden2002}.  The inferred distribution for $l/R$ obtained by \citet{asensio2013} was very broad and roughly all values of $l/R \in [0,2]$ are possible, with the maximum of the distribution around $l/R \approx 0.8$. Although the large values of $l/R$ obtained by \citet{verwichte2013} and \citet{asensio2013} clearly violate the assumption $l/R \ll 1$, the error associated with the use of the  TTTB approximation was not taken into account. In our view, in those studies the use of the full numerical inversion scheme would have been more appropriate than the analytic TTTB scheme. Indeed, the possible future use of the more robust numerical scheme  is mentioned by \citet{verwichte2013}. In addition, both \citet{verwichte2013} and \citet{asensio2013} restricted themselves to a specific variation of density in the transitional layer, namely a sinusoidal profile, and ignored that the inferred seismic intervals were directly influenced by that specific choice of density variation. These two facts may add some uncertainty to the seismic intervals obtained in both articles, specially for the strongly damped events, and should be considered in forthcoming studies.

Seismology schemes for coronal loop transverse oscillations have also been adapted to the case of transverse oscillations of thin threads in solar prominences \citep[e.g.,][]{soler2010,arreguiballester2011} since equivalent theoretical models are used in both cases. As for coronal loops, little is known about the spatial variation of plasma density across the thin prominence threads. Hence, the results of this paper are also relevant for the inversion of physical parameters in prominence threads.

Finally, we would like to mention  the potentiality of the  Bayesian methods used by \citet{arregui2011baye} and \citet{asensio2013}. These methods have been shown to be very useful to perform model comparison  and to determine the plausibility of each possible model \citep[see][]{arregui2013baye,arregui2013prop}. The ability of Bayesian analysis to distinguish between different transverse density profiles should be exploited in future works.

\acknowledgements{
We are very grateful to I. Arregui for giving valuable comments. We also thank J. Andries for his remark about the importance of determining the actual resonance position for the parabolic profile. The authors acknowledge support from MINECO and FEDER funds through project AYA2011-22846, and from CAIB through the `Grups Competitius' program and FEDER funds. M.G. acknowledges support from KU Leuven via GOA/2009-009. The research of M.G. has partially been funded by the Interuniversity Attraction Poles Programme initiated by the Belgian Science Policy Office (IAP P7/08 CHARM). J.T. acknowledges support from MINECO through a Ram\'on y Cajal grant.}

\bibliographystyle{apj} 
\bibliography{refs}

\end{document}